# Process signature-driven high spatio-temporal resolution alignment of multimodal data

Abhishek Hanchate[a], Himanshu Balhara[a], Vishal S. Chindepalli[a], Satish T.S. Bukkapatnam[a*]

[a]Wm Michael Barnes'64 Department of Industrial and Systems Engineering, Texas A&M University, 3131 TAMU, College Station, Texas 77840, United States

**ABSTRACT**

We present HiRA-Pro, a novel procedure to align, at high spatio-temporal resolutions, multimodal signals from real-world processes and systems that exhibit diverse transient, nonlinear stochastic dynamics, such as manufacturing machines. It is based on discerning and synchronizing the process signatures of salient kinematic and dynamic events in these disparate signals. HiRA-Pro addresses the challenge of aligning data with sub-millisecond phenomena, where traditional timestamp, external trigger, or clock-based alignment methods fall short. The effectiveness of HiRA-Pro is demonstrated in a smart manufacturing context, where it aligns data from 13+ channels acquired during 3D-printing and milling operations on an Optomec-LENS® MTS 500 hybrid machine. The aligned data is then voxelized to generate 0.25 second aligned data chunks that correspond to physical voxels on the produced part. The superiority of HiRA-Pro is further showcased through case studies in additive manufacturing, demonstrating improved machine learning-based predictive performance due to precise multimodal data alignment. Specifically, testing classification accuracies improved by almost 35% with the application of HiRA-Pro, even with limited data, allowing for precise localization of artifacts. The paper also provides a comprehensive discussion on the proposed method, its applications, and comparative qualitative analysis with a few other alignment methods. HiRA-Pro achieves temporal-spatial resolutions of 10-1000 $\mu$s and 100 $\mu$m in order to generate datasets that register with physical voxels on the 3D-printed and milled part. These resolutions are at least an order of magnitude finer than the existing alignment methods that employ individual timestamps, statistical correlations, or common clocks, which achieve precision of hundreds of milliseconds.

## 1 Introduction

Modern manufacturing systems are increasingly becoming instrumented with a variety of sensors. The use of internet-of-things (IoT) sensors is projected to grow at a 40% rate annually [1, 2], leading to an increased availability of diverse multimodal sensor data. In this scenario, different data streams such as those capturing machine vibrations [3, 4], acoustic signatures [5, 6], temperature [7, 8], etc., are collected at different frequency rates. Machine communication and control data streams, as well as the set points of various supervisory elements, all occur at different sampling rates, latencies, and are captured on diverse data acquisition systems (DAQs). They are digitized and handled differently. Although combining information from multiple (multimodal) sensors is recognized to greatly improve our capacity to monitor and control the process [9-12], drawing conclusions from unaligned sensor data poses considerable risks that should not be overlooked. Data alignment is a growing facet with the increasing proliferation of multimodal data with various applications, yet it is currently being ignored.

The issue of alignment is similar to that of the parable, "blind men and an elephant" [13], which is a story of a group of blind men who came across an elephant for the first time in their lives. Each blind man touches a different part of the elephant's body and then biasedly describes it according to their limited experience as shown in Fig. 1(a). In case of multimodal data alignment, each blind man would represent a sensor and talk in a different language. There is not one elephant, but it is a rapidly changing and nonlinear dynamic process [14-16] with many different animals streaming through a lane. During the streaming, whatever may happen at one timestamp might differ from other timestamps. The thought that these blind men can infer accurate information together at the same time is the classic idea of sensor fusion. For accurate sensor fusion and effective inference, all sensors must touch the same aspect of the process at the same time, and this corresponds to alignment. Figure 1(b) provides an illustration of this analogous relation of multimodal data alignment with the described parable. It consists of three different stages of the process chain, each of which is comprised of monitoring based on high speed camera footage, vibration and acoustic emissions signals across the multimodal data axis. The process itself is also depicted in Fig. 1(b). in terms of 3D-printing, milling, and precision milling along the direction of the process chain axis. Figure 1(c) illustrates the machine and experimental setup employed in this work in terms of an Optomec MTS 500 hybrid machine and annotated depiction of some of the multimodal sensors. Our work involves aligning multimodal data at a voxel level of the part as it is being made via different stages of a process chain as shown in Fig. 1(d).

The moral of this alignment problem and the parable is that that humans/sensors tend to claim absolute truth based on their limited, subjective experience while ignoring other people's/sensors limited, subjective experiences which may be equally true. With the exponential growth of multimodal data and quality assurance, the aspect of multimodal alignment has a lot of importance and is potentially very impactful.

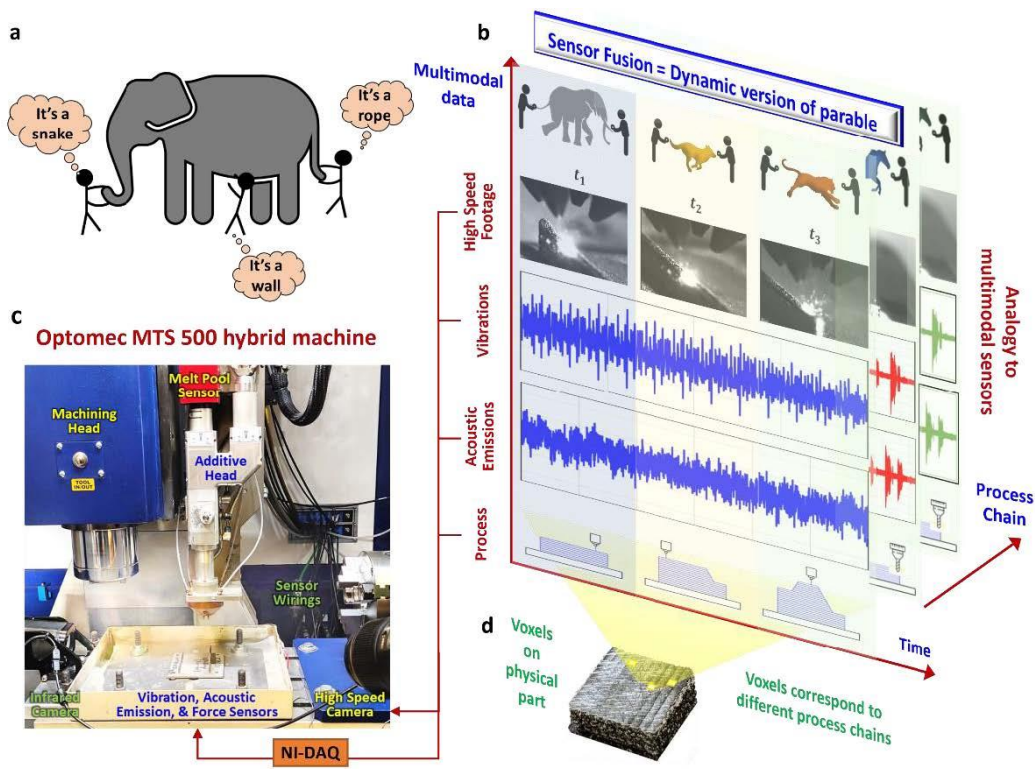

Figure 1. a) The parable of the blind men and an elephant, wherein three men describe three different body parts of an elephant with their limited subjective experience b) An analogy to dynamic version of the parable in terms of multimodal sensors (video frame set and signals from an accelerometer and acoustic emissions sensor) instrumented to monitor dynamic manufacturing process stages such as 3D-printing, milling, and precision milling c) Optomec MTS 500 hybrid machine with some of its sensing capabilities d) A voxel (3D pixel) on the physical part

The importance of data accuracy, consistency, and privacy has spiked up tremendously in recent years [17-20]. The National Institute of Standards and Technology (NIST) hosted a small-scale workshop in 2015 to evaluate heterogenous data from traffic and weather sensors that are typically collected in highway traffic domain [21]. The focus was on addressing major challenges and enhancements necessary to drive the application of data science forward. One of those challenges was that of data synchronization and alignment in the case of multimodal sensor data. Synchronized and aligned data is crucial in order to harness the full ability of an algorithm or model, as it improves the efficiency with which computers retrieve and allocate memory [22]. In the highway traffic domain, for example, data between the traffic camera and the accident reports will need alignment in order to retrieve the timestamps in which the accidents occurred in order to further observe the reason behind a specific accident.

Issues that might seem like minor errors in the data can have adverse negative effects on the performance of machine learning (ML) models and lead to biased decision making [23, 24]. Moreover, the issue of unaligned data gets worse as the number of data streams increases. Data alignment issues arise due to unsynchronized and improper collection of various sensor data sources and specifications of DAQs, which can adversely affect the quality of the collected multi-channel data [25]. With the high rates of data collection in smart manufacturing processes and setups, these issues tend to amplify with increasing amounts of non-synchronous times and DAQ clock delays [26, 27]. For example, a typical Acoustic Emissions (AE) sensor operates in the frequency range of few hundreds of kHz or MHz [6]. In such cases, data synchronization and alignment become essential, especially to make accurate and timely millisecond and microsecond decisions.

Data alignment involves aligning data events by time or in the order they occur in. It is the process of synchronizing data pertaining to two or more devices and updating differences between them to provide consistency within DAQs. Such alignment enables and assures congruency among all the sources of data and keeps the acquisition consistent. Multimodal aligned data can then be segmentized in a way that small segments of aligned data correspond to a small part (voxel) of the process physically.

In this work, we propose a novel algorithmic approach for high-resolution alignment of multimodal spatio-temporal data called HiRA-Pro and are looking to bring out the important facet of process physics-driven alignment. Process signatures are derived from the underutilized aspect of process kinematics and dynamics. Every process has multiple salient kinematic and dynamic events. These usually show up as visible clues in the data acquired from commercial-off-the-shelf (COTS) or image-based sensors, that can capture many aspects of the underlying physics. HiRA-Pro uses them to derive process signatures that can be then used as markers to drive multimodal data alignment. This is especially promising in a manufacturing context since the majority of process physics involves sub-millisecond time phenomena wherein statistical correlation-based alignment does not suffice. While timestamp or clock-based alignment can provide decent resolutions, the requirement for milliseconds scale of alignment makes them insufficient. Hence, necessitating the need for process signatures to be brought in the alignment process for obtaining accurate results. In a way, HiRA-Pro is registering the data signals back to the physical phenomenon to align them better.

The effectiveness of the proposed HiRA-Pro approach is demonstrated from a smart manufacturing implementation point of view. Myriad COTS sensors are employed via a smart sensor-wrapper and sensing suite to acquire multimodal data during 3D-printing and milling operations on an Optomec-LENS® MTS 500 hybrid machine. In total, data pool consists of 13+ channels, namely those acquired via an accelerometer, AE sensor, thin-film piezoelectric sensor, thermal meltpool sensor, high-speed camera, smartphone camera, and a data logger that keeps track of 7+ machine and controller-based variables. The multimodal data was aligned via HiRA-Pro, achieving sub-millisecond resolutions, and later voxelized (segmentized) to generate 0.25 second aligned data chunks that correspond to physical voxels (segments) on the produced part. Every signal in these voxels then corresponds to an output of a particular physical phenomena in the process. The data was also aligned via time-based multimodal alignment and the details have been covered in the supplementary section.

The commercialization value and need for such algorithmic data alignment approach was also evaluated as a part of the National Science Foundation (NSF) Texas A&M University I-Corps site program. The interviews suggest a significant requirement in the industry for such automated means, especially in process industries such as Intel.

The remainder of the paper is structured as follows - Section 2 compiles relevant background with review of existing literature and gaps. Existing methods for alignment and their usecases have been discussed along with related research topics like data registration and voxelization. The alignment problem has been mathematically described in section 3 with a formal state-observer formulation. The details and algorithms corresponding to the proposed process signature-based alignment (HiRA-Pro) are covered in section 4. Section 5 demonstrates the superiority of HiRA-Pro by applying it to one specific and another broader case study in the context of additive manufacturing. The specific case study showcases improved ML-based predictive performance of porosity in a hybrid Directed Energy Deposition (DED) process due to multimodal data alignment while the broader one provides a holistic view and general results with many more modalities. The experimental setup and description of the multimodal sensor data extracted during DED-based 3D-printing and milling operations on the Optomec-LENS® MTS 500 hybrid machine is also covered. Section 5 also covers some discussion in a comparative manner among major multimodal alignment methods and the paper concludes with a few concluding remarks and laying down of future work in section 6. The supplementary section covers algorithms and description of time-based multimodal data alignment, that is proposed as a quick and efficient means of achieving alignment based on timestamps and DAQ clocks.

## 2 Background and literature review

### 2.1 Multimodal data and its classification

In a manufacturing setting, multimodal data refers to the integration of diverse types of sensor data such as time series-based, image- or intensity-based, and machine log-based data, among others. Such data is collected at various stages throughout the manufacturing process and can provide a comprehensive understanding of the production workflow. The fusion of these disparate data modalities can enhance the decision-making process, enabling more accurate predictions and comprehensive insights. For instance, data from machine vibrations can be combined with image data from visual inspections to improve defect detection. Similarly, log-based data can be used alongside other sensor data to optimize process efficiency and reduce downtime. The aspect of multimodal data plays a crucial role in enhancing the intelligence and efficiency of manufacturing systems, triggering the need for fully utilizing it. Such multimodal data can be classified mainly into the categories described in sub-sections 2.1.1 through 2.1.3.

#### *2.1.1 Time series-based data streams*

Most sensors are designed to measure some physical or environmental variable over a period of time and capture the measurements in form of sequential data taken at regular or irregular intervals. Such sensors can be referred to as time series sensors and typically consist of a sensing element that is responsible for converting the physical variable measurements into an electrical signal, typically in voltage or current. This signal is then conditioned and amplified before being captured and stored by a DAQ. The sensing element could be a strain gauge, pressure transducer, or a thermocouple, depending on the physical variable to be captured.

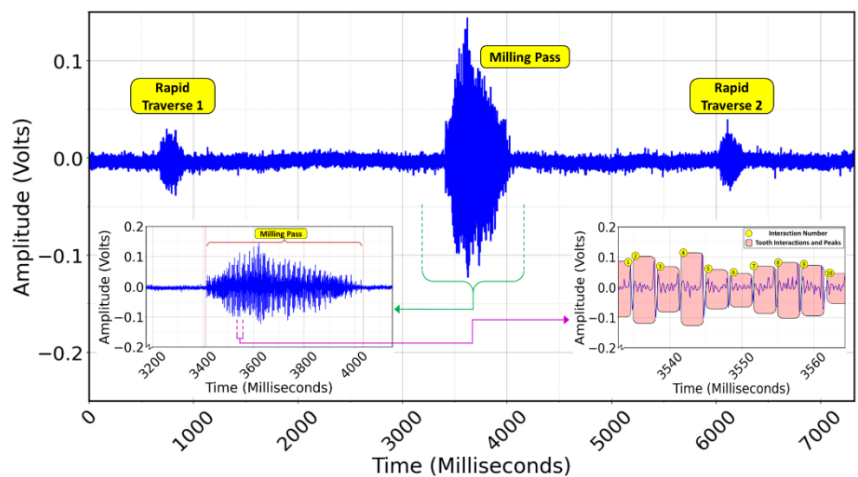

**Figure 2. Time series-based data signal acquired during a manufacturing milling operation**

Examples of such sensors include vibrations captured via an accelerometer or forces captured by a piezoelectric thin-film sensor. Figure 2 depicts a typical time series-based data stream or signal captured by such sensors. The physical quantity (vibration) being measured is converted to voltage as shown on the y-axis and the sensor captures variations of this voltage across time at a rate of 8192 Hz as the process continues. The signal in Fig. 2 corresponds to a part of a signal captured during a upmilling process of a 3D-printed part with feed rate of 4.23 mm/sec and a spindle RPM of 3000 with a milling tool with 6 teeth. The dimension of the part being milled is approximately 10 mm. In total, there are at least 3 salient process events easily differentiable from rest of the baseline signal. The smaller amplitude bursts are captured during rapid traverses of the table and the larger one corresponds to a milling pass. Considering only the milling pass, the feed per tooth would be $4.23/(3000/60 \times 6) = 0.0141$ mm and each tooth interacts with the workpiece for $0.0141/4.23 \approx 0.0033$ seconds, which in turn translates to $0.0033 \times 8192 \approx 27.306$ samples in the signal. Time series-based signals possess the ability to capture large amounts of time-varying data and can capture the underlying physics behind the process to a great extent, making them viable for state-of-the-art ML modeling techniques. In fact, the information on the tooth interactions is captured in the time series depicted in Fig. 2. By zooming in further within the milling pass (left subfigure in Fig. 2), it can be noted that it approximately takes 600 milliseconds, which is lower than expected since the tool did not always interact with the part (due to unfinished 3D-printed top surface). However, within the pass itself, further zooming in, the tooth interactions are unraveled as shown in right subfigure in Fig. 2. Approximately 10 interactions are captured over a span of about 30 milliseconds or 260 samples. The patterns observed within each interacted as indicated on right subfigure of Fig. 2 are consistent with domain expertise wherein upmilling would typically result in a short spike, followed by an increase over time, and finish with a larger spike. This is also consistent with the time for individual tooth interaction, i.e. 27.306 samples, which becomes 273.06 over 10 tooth interaction events. This makes such time varying signals to be excellent candidates for achieving an alignment based on process signatures. Depending on the desired resolutions, it is possible to use these tooth interaction dynamic signatures as the driving markers for the alignment.

*2.1.2 Intensity-based data streams*

With the onset of smart sensors and advanced camera videography tools, it is possible to capture beyond just time-varying data in the form of a time series. Digital image-based sensors or devices can capture image frames and convert them into digital data for further analyses. These image-based sensors play a pivotal role in modern manufacturing and convert light (photons or intensity values) into electrical signals as well as possess the capability for raw image frame collection. We refer to such data as intensity-based data streams. Such devices can be based on charge-coupled devices (CCDs) such as common digital cameras or medical imaging devices or on complementary metal-oxide-semiconductors (CMOS) such as modern smartphones and drone cameras. They can also be classified based on chroma and shutter type in the device, frame rate, resolution, and pixel size.

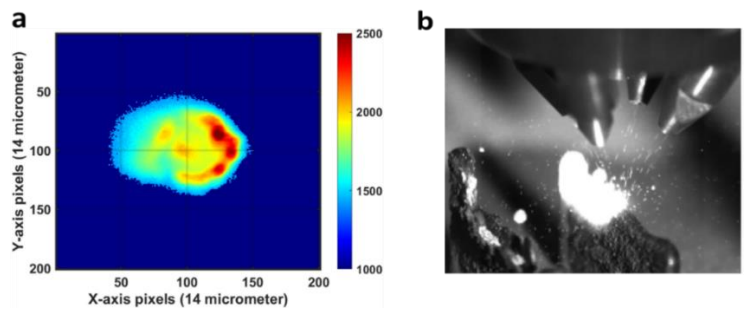

**Figure 3. Example of image-based data streams - a) thermogram (melt pool image) that shows thermal variations during a 3D-printing process b) high speed camera frame (at 6000 FPS) capturing a high frame rate video during 3D-printing**

The intensities captured can range from video camera clues to any calibrated form of capture such as temperature gradients in form of light intensities as shown in Fig. 3(a). Such calibration is typically done for thermal or melt pool-based sensors. Another typical image-based sensor involves the use of high speed cameras that allow for capturing high speed and high resolution processes with capabilities for slow motion analysis of the frames. The frame rates can be in the order of several hundred thousand frames per second, but is only capable of capturing frames for equivalently smaller durations of time. Figure 3(b) illustrates a frame captured via a high-speed camera at 6000 frames per second. Such intensity-based data streams can successfully capture process signatures on their frames, but the derivative information based on these frames can provide additional insights. For instance, Fig. 4 depicts such a derivative in the form of average intensities captured across frames for a cropped region of interest. The source video corresponding to this was a smartphone camera footage capturing a 3D-printing process, focusing on an individual print layer with 9 printing tracks and an initial perimeter according to the desired part dimension. It is clear from Fig. 4 that intensities capture information about these tracks and the perimeter in the form of intensity spikes and drops. This is because the laser is on and off between each of these process events. The aspect of laser turning on and off provides us with process signatures that can be utilized to align intensity-based data streams with other modalities and with the process itself.

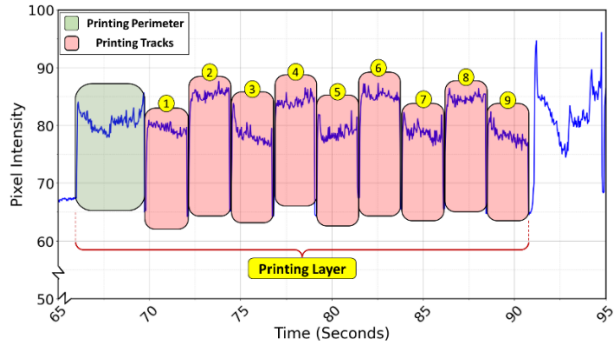

**Figure 4. Derivative information in the form of average pixel intensity across frames for an optical video footage captured during part of a 3D-printing process**

*2.1.3 Machine log-based data streams*

Modern CNC machines are able to capture logs of the machine component positions, status of ON/OFF components, and any form of event triggering. For example, Open Platform Communications (OPC) Unified Architecture (UA) data loggers can be used to capture the motion of a table within a CNC in the form of movement in the $x$-, $y$-, and $z$-direction as shown in Fig. 5. The first column elements, 1 or 2, correspond to the $x$ and $y$ direction motions respectively, second column quantifies this, and third column timestamps the same. Depending on the setup, it is possible to capture hundreds and thousands of other such variables. Moreover, such an event log can be considered as a table of multiple data streams, each represented by a column.

| 1 | 0.015 | 2022-07-26T17:29:00.802Z |
|---|---|---|
| 2 | 0.029444882 | 2022-07-26T17:29:00.904Z |
| 2 | 0.056644882 | 2022-07-26T17:29:01.108Z |
| 2 | 0.097444882 | 2022-07-26T17:29:01.414Z |
| 2 | 0.124644882 | 2022-07-26T17:29:01.618Z |
| 2 | 0.165444882 | 2022-07-26T17:29:01.924Z |
| 2 | 0.192644882 | 2022-07-26T17:29:02.128Z |

**Figure 5. Example of a machine log-based data stream capturing two variables (1 and 2) in column one with their corresponding values in column two and timestamps (UTC format) in column three**

The multimodal data stream classification is not limited to the aforementioned categories but provides for a brief categorical allotment. The synchronization and alignment methodology can differ depending on the category to which a given sensor stream belongs from such classification. The algorithms proposed in this work take this classification into account to exploit the array of techniques to achieve aligned multimodal data.

## 2.2 The aspect of multimodal data voxelization and multimodal machine learning

The concept of voxels has existed in literature relevant to 3D computer graphics, computer vision, and computational biology since the mid 1900s. Typically, voxelization refers to the process of converting continuous geometric representations into discrete volumetric data wherein the volume is subdivided into small and regularly spaced cubic units known as voxels [28]. These regular grids of voxels can be thought of as 3D counterparts to pixels in 2D images. They contain information like material properties and other application dependent attributes, and can be used for tasks of part optimization, simulations, and visualization [29]. Another aspect of voxelization emerged recently in the realm of manufacturing wherein studies have focused on voxel printing in additive manufacturing by 3D printing parts and assembling printed voxels layer-by-layer [30-32]. This has led to possibilities for voxel-wise control over 3D-printed part's volume, allowing for parameter tuning during the process and resulting in multifunctionality in the final product [32]. Very few works exist that include the multimodal data aspect into this voxelization, especially in the manufacturing domain. Assuming a similar analogy, data voxelization would involve converting continuous data signals into discrete or continuous regular grids of the data that correspond to voxels (3D pixels).

In the proposed work, we extend the voxelization concept to include aligned multimodal data with diverse modalities such as videographic data, time-series based signals, and event logs along with the process and product information. The resulting voxels thereby represent segmentized aligned multimodal data that corresponds to a grid of 2D/3D-space on the manufactured part and capture the relevant process information. However, we relax the requirement for regular grid nature of these voxels, instead considering cuboidal over cubical voxels. This is somewhat analogous to the data registration problem, however since alignment can be a part of it, we refer it as multimodal data voxelization.

Depending on the desired resolution of alignment, the proposed data voxelization can be implemented after or during the multimodal data alignment procedure. In this work, we only demonstrate the former and expect even better alignment with the latter. The aligned data can be voxelized to extract portions corresponding to small segments of the process (for instance, only first track of the second layer of a 3D-printing process), resulting in aligned multimodal data voxels of desired dimensions. Such voxels can be extracted over the entire span of the process as required. By doing so, it is possible to obtain a discretized representation of the process while linking various modalities of sensor data. These representations, especially in context of additive manufacturing, can be easily translated to tool paths, making it viable to generate layer-by-layer or track-by-track instructions and perform subsequent analyses. This can allow for precise control of material deposition and other printing parameters due to the high-resolution communication between multimodal data and the process itself.

Such data voxelization also allows for mapping the aligned multimodal data to various aspects of final part's characterization. The case studies in section 5 cover a short discussion on the same in case of DED-based 3D-printing process wherein process parameters and surface profiles of the 3D-printed part via optical microscopy and profilometry are involved in the alignment and voxelization procedure. Such multimodal data voxelization promotes efficient processing and further analyses via ML models given the grid-like form. While most ML techniques can still perform adequately with unaligned and low-quality data without grasping the complete power of multimodal nature of sensing, it is essential that they interpret multimodal data together at the same time to understand the process completely. Multimodal ML is a thriving topic of research with excellent potential but some of the challenges have not been studied extensively in literature. Multimodal data alignment is one such issue wherein to truly extract all the available information with no noise or false feature-based ML applications, it is extremely important to align and synchronize all the available data signals to milli- and micro-second resolutions. The next three sub-sections provide some literature review relevant to the alignment problem.

## 2.3 Issues with data acquisition pipelines and time-based multimodal data alignment

Modern day sensors and imaging systems provide some means of accessing associated time information, either directly or indirectly. Most sensors and modern sensing suites come with built-in capabilities to capture timestamps associated with the signal to be captured. For example, National Instrument (NI) DAQs come with a built-in clock and are able to capture signal amplitudes in a time series format. Usually, these clocks capture time information in the form of timestamps, which refer to the time at which a particular reading was taken on a sensor or when a sensor detected an event occurring. The timestamps can be set according to the desired format and resolution by modifying the collection sampling rate of the DAQ. For example, a 10 kHz sampling rate will capture 10000 data points in form of signal amplitudes and their corresponding timestamps with a resolution of 0.1 ms.

Depending on the sensor manufacturer or the data acquisition methodology, the definition of a timestamp might vary. For example, sophisticated high-speed imaging systems come with IRIG timestamps that correspond to the real time at which a given frame is captured. On the other hand, in some COTS sensors which require separate data acquisition systems require digitization and might relay this information as the timestamps instead of when the sensor picked up something. This delay between when a sensor picks up a signal and when the corresponding data is acquired and timestamped can vary depending on a myriad of factors such as signal conditioning, processing time, and communication latency between the sensor and the DAQs. Signal conditioning usually involves preparing the analog signal that is picked up by the sensor for digitization by the DAQ. Such preparation includes signal amplification as well as filtering among others that introduce time delay in the overall system. Processing time corresponds to the time required for the digitization and depends on factors such as the speed of the analog-to-digital converter (ADC), the system's overall processing power, and the complexities of the algorithms involved. Communication latency, which refers to the time required for transmission of the signal from the sensor to the DAQ and from the DAQ to the storage device or system computer, is a common issue in today's data-driven world. Such latency depends on a variety of factors such as the communication protocol that is implemented and the availability of appropriate bandwidth. Overall, this time delay can range from a few microseconds to even several milliseconds or more depending on the specifications of the system.

Due to the minute-scale nature of this delay and little-to-no concern by those who have employed a multimodal sensing-based implementations, not many efforts are present in current literature to address high resolution time-space alignment of the data channels. Prior works tend to mimic multimodal data alignment as a preprocessing step while fusing data from various sensors before using the fused dataset for ML modeling. Basic ML and deep learning models still work based on the hidden patterns discovered in the data, but there is always a scope for further improvement of performance by achieving better alignment resolutions.

In this paper, we also provide details about our own take on a time-based multimodal data alignment approach (see supplemental material section S.1 and corresponding algorithms S0 and S1), which provides means of accessing and extracting time information for various types of data streams. Benchmarking and testing these algorithms on various data streams indicate results that correspond to time-based alignment with millisecond resolutions. The results can vary depending on factors such as acquisition rates of data streams and selection of baseline for clock synchronization.

While time-based multimodal data alignment (see supplemental material section S.1) provides excellent resolution results, it also comes with a few challenges. First, clock delays can be very random in nature, making it difficult to align various data streams due to the requirement for proper synchronization of clocks. To further amplify this issue, these clocks require maintenance without which the time can quickly go out of sync again. An example of this is laptops that require clock synchronization from time to time. An example in the manufacturing domain are controllers such as the Siemens 828D controller, that are involved in timestamping of various events and changes in machine log-based data streams. These controllers can lag by 2 minutes or so every 15 days if not always plugged in to the internet. However, such connectivity requirements are often difficult or undesirable in practical settings, especially with the onset of 5G and bloom in potential cyber-attacks. Moreover, the variability in the time zones, units, and formats of timestamps, as well as other internal delays can create difficulties in generalizability of such time-based alignment across a wide range of sensorized processes and setups.

Another issue that arises is in the form of proprietary restrictions that do not allow extraction of time information from data streams. In such cases, time-based multimodal data alignment is not possible without excluding the data stream of issue from the alignment methodology. Additionally, as discussed earlier, the timestamps do not necessarily represent a particular event accurately in time due to inconsistencies arising from time taken for digitization and conditioning of signals, as well as latency issues. To address these issues, we propose another set of algorithms that do not rely on timestamp information and instead derives a process signature-based multimodal data alignment.

## 2.4 Data alignment and synchronization during acquisition

A methodology for acquiring somewhat aligned datasets involves hard or physical triggering systems that use hardware components based on specific physical events to trigger the acquisition of multimodal data. These systems gained popularity due to modern sensing capabilities, enabling the acquisition of vast amounts of data, reaching several hundreds of millions and billions of samples with a combined volume in terabytes. In real-time execution, such acquisitions can lead to data volumes several orders of magnitude higher. For example, experiments at CERN's particle accelerator have combined data volumes exceeding 60 million megabytes per second, equivalent to over 5400 simultaneous 4k video streams [33]. However, in many scenarios, only a subset of the captured data contains events with distinguishing and interesting characteristics for further analyses [34]. This is also applicable to fields such as manufacturing, which involve high rates of data acquisition. Such selective data acquisition techniques offer another approach to data synchronization and alignment [35, 36]. CERN employs this method in their large ion collider experiment (ATLAS) using the ATLAS trigger system [37], which employs a special event selection methodology (trigger) to pick events useful for further analysis.

In the field of 3D scanning, object tracking, robotics, and augmented reality (AR), structured light systems (SLS) project patterns of lights onto a scene to capture depth information [38]. While their main application is not data triggering, SLSs can be repurposed to trigger data acquisition under specific conditions, optimizing the process and reducing the amount of data to be processed. Triggers can be pattern-based [39], where a specific pattern is projected to initiate data acquisition, or they can rely on changes in the temporal properties of the light pattern, such as object movements or surface reflectivity changes [40]. Additionally, triggers can be based on detecting specific geometric features like corners or edges [41]. Huang *et al.* [42] utilized SLS with an event camera to capture pixel-level intensity changes asynchronously with high temporal resolution and motion blur reduction. Liu *et al.* [43] emphasized the importance of time synchronization in autonomous systems and achieved precise circuit-based timestamping for cameras and inertial measurement units.

Analogous to the SLS-based methods in 3D scanning, the manufacturing domain employs hardware-based or physical triggers for selective data acquisition. These triggers use physical events, like mechanical or light intensity changes, to initiate simultaneous data acquisition from multiple sensors, ensuring precise time alignment of all data streams. Such triggered acquisition improves the quality of datasets for various manufacturing applications, including real-time monitoring and process control.

However, implementing hard triggering elements introduces additional costs to the sensorization system, becoming an integral part of the setup with the need for frequent and accurate calibration, leading to undesirable downtime costs. Moreover, due to the diverse sensor modalities, these hardware-based triggers may not be readily compatible with all sensors, often requiring substantial additional hardware. Even if implemented, practical challenges arise, such as unpredictability and randomness caused by varying lighting conditions, scalability limitations, and sensitivity issues with hardware systems. Such triggered acquisition also limits the acquisition to only parts of the process, which is undesirable in some situations. Multimodal data alignment done during the acquisition is not applicable to existing data that might have been acquired without the necessary hardware. So, there is a need for post-acquisition means of achieving such a feat. The next sub-section addresses existing works of post-acquisition multimodal alignment, synchronization, and registration, a broader problem involving spatial alignment, in detail.

## 2.5 Data alignment and synchronization post-acquisition

Oftentimes in literature, data alignment and data synchronization are considered synonymous. Recently, there's been a growing interest in synchronizing multiple sensors' signals, especially those embedded in wearable devices. Bennett *et al.* [44, 45] have explored various methods for synchronizing multi-sensor data, leveraging physical and cyber coupling between sensor data streams. The authors aligned physical events, sensor data, and clock accuracy estimates, and achieved a significant 60% reduction in average system drift, from 495.8 ms to 194.8 ms [46].

Cippitelli *et al.* [47] demonstrated the synchronization of heterogeneous sensors, such as RGB-Depth cameras and wearable sensors. They used a test bed with a PC connected to a camera and wireless sensor, achieving synchronization by tracking camera activities with a delay and using a LED as a reference. In a related study, Zhang *et al.* [48] tackled synchronization issues between video cameras and wearable sensors. Their novel approach, called Window Induced Shift Estimation (SyncWISE), utilized a wearable camera and an accelerometer, achieving an impressive synchronization accuracy of 90% with a tolerance of 700 ms.

Huck *et al.* [49] proposed a synchronization approach wherein they developed a filter-based method for achieving accurate temporal synchronization by removing jitter and minimizing time delays in measurements. Shaabana and Zheng [50] introduced CRONOS, a data-driven framework for sensor data synchronization in wearables and IoT devices, which employs a graph-based approach. CRONOS extracts common events between sensor streams for synchronization and demonstrated a remarkable 98% improvement in system drift, with an error of only 6 ms at a 100 Hz sampling rate. Wang *et al.* [51] proposed a hybrid synchronization approach that combines Network Time Protocol (NTP) and physical events. This hybrid approach was shown to achieve more precise synchronization than standalone techniques, reducing the error to 20 ms over 15 hours.

There has been limited work pertaining to the synchronization and alignment issue in the manufacturing domain. Feng *et al*. [52] proposed a general data alignment procedure for laser powder bed fusion methods wherein multimodal data streams consisted of melt pool images, scan paths, layer images, ex-situ X-ray computer tomography (XCT) 3D model, coordinate measurements, and a 3D computer-aided design (CAD) model. The authors established a good foundation and provided initial guidance, but the work was only a conceptual exploration. It assumes that data clocks are already in-sync and does not consider any time series-based data streams. The authors recommend that manufacturing problems like detection of defects, anomalies, and cause analysis should be done via aligned multimodal data. Other similar prior works include spatially aligning data streams to a single coordinate system and format.

A broader version of the data alignment problem is that of data registration [53] wherein both temporal and spatial alignment of multimodal data streams are required. Kim *et al.* [54] proposed a deep learning-based approach for registering and aligning meltpool images in laser powder bed fusion-based additive manufacturing. Deep convolutional neural network was utilized to extract image features, and the images were registered by a transformation model based on the features. Feng *et al.* [55] proposed a methodology for registering geometric data acquired both via in-situ and ex-situ during additive manufacturing process. The work focuses on the importance and need of accurate registration and their methodology results in data registration with average error of 0.005 mm. Lu *et al.* [56] proposed methodology for camera-based coaxial melt pool monitoring (MPM) data registration based on the coordinate system for the build volume. Only melt pool images and related process information was considered to obtain alignment between the images and real laser positions by removing delay during calibration process to achieve a positional error of less than 1% for all the frames.

Despite significant advancements in sensor data synchronization, alignment, and registration techniques, current approaches primarily cater to sensors of similar or limited modalities, which restricts their applicability. Furthermore, to fully harness the capabilities of modern sensors, extremely high sampling rates are essential. However, most existing studies limit themselves to sampling rates of only a few hundred to a few thousand Hertz. Additionally, there is a need for more versatile calibration or setup procedures, as current methods may not be flexible or adaptable enough to cater to different scenarios or requirements. Many existing studies provide mere guidance without any real-world examples or use-cases, making it difficult for practitioners to apply these techniques in practical scenarios.

Almost every process involves underlying physics phenomena that unfolds on a sub-millisecond timescale. Traditional statistical correlation-based means of alignment often do not suffice in capturing these rapids events. The underlying process physics often goes underutilized, indicating a need for techniques that can effectively capture and utilize the process physics to enhance the accuracy and reliability of the results. Addressing these gaps could significantly contribute to the field and enhance the applicability and effectiveness of sensor data synchronization, alignment, and registration techniques. The process physics signatures need to be leveraged in multimodal data alignment for developing methods that are applicable to most scenarios, are generalizable, and provide accurate and high resolution aligned datasets.

The methodology in the proposed work addresses these literature gaps and limitations by providing such process physics-driven algorithms that can be used depending on desired resolution needs of alignment. The validation and testing are carried out on real-world use cases that involve sensor data of varying modalities ranging from time series-based signals that are sampled at very high rates, intensity-based sensor data such as melt pool (thermograms), high speed camera videos and optical video captures, as well as other log-based data. The algorithms may also allow for including data that captures characterization and process parameters in the form of CAD models, XCT images, process information, and material microstructures to a certain extent. This is partially demonstrated in this work in case of data from optical microscopy and profilometer images as a part of case study in section 5. Section 3 will introduce this multimodal alignment problem as a mathematical treatment.

## 3 Formulation of the alignment problem

This section covers a formal description of the problem in terms of a state-observer structure [57]. In dynamic systems theory, a state-observer form is used to estimate unmeasured state variables of a system. The observer relies on mathematical equations and available measurements in order to estimate the system's internal state. Consider a process where multimodal data is collected by various sensors working together as shown in Fig. 6. Assuming that the process is deterministic (no noise is present), the state space corresponding to it can generally be represented as:

$$\dot{x} = f(x(t)) \tag{1}$$

wherein state space $\dot{x}$ depends on thermo-mechanics involved in the process. The multimodal sensor data such as those shown in Fig. 6. can be considered as output from this process. More specifically, each of the sensor data captures a particular observable subset (subspace) of the process. The outputs are in the form of discretely sampled data:

$$y_{1k}, y_{2k}, \ldots, y_{ik} \tag{2}$$
$$k = 1, 2, 3, \ldots, n_i$$
$$i = 1, 2, 3, \ldots, m$$

where $m$ is total number of data outputs or streams, $i$ corresponds to the data output index, and $k$ corresponds to the sample index in the discretized data output. For instance, the sample data $y_{ik}$ will correspond to the $i$th data channel and $k$th sample index. It is important to note that each channel may have a different number of sample indices $k$ since they are extracted at different sampling rates. Each of these $y_{ik}$s are nothing but some sort of $y_i(k)$. For a fixed sample index $k$, the $i$th channel's $y_{ik}$ might be realized at time $t_{ik}$ while the $j$th channel's $y_{jk}$ might have been obtained at $t_{jk}$. Therefore, $y_{ik}$ might not capture the same thing as $y_{jk}$. This is where the misalignment happens!

Each process is represented by its state space and within the state space the process executes a particular trajectory. For simplicity, assume that the process as well as observations are noise free. Such a process has different outputs, and each sensor's data could be a particular function of process's state. Sensor channel $y_{1k}$ could be a particular function of the state space $\dot{x}$ such that:

$$y_1(k) = h_1(x(t_{1k})) \tag{3}$$

Similarly, $y_{2k}$ could be some other function of the state space $\dot{x}$:

$$y_2(k) = h_2(x(t_{2k})) \tag{4}$$

and so on, where $h_i$s are some functions of the state space. Both, the function as well as the time might differ from $y_{1k}$ to $y_{2k}$ as illustrated in the schema in Fig. 6. We have different states and outputs, and such a representation is known as state observer form that incorporates the process and its outputs. From the outputs, one can reconstruct an observable subset of the process. This is equivalent to observing a particular trajectory of a process or particular projection of the process on a manifold. Let the reconstructed state space be $\Phi_i(t_{ik})$, a function of $y_i(k)$ such that:

$$\Phi_i(t_{ik}) = g_i(y_i(k)) \tag{5}$$

where $g_i s$ are some function of $y_i(k)$, $i$ are data output indices, and $k$ corresponds to sample indices in discretized outputs. From $y_1(k)$, you will get particular trajectory in the subspace $\Phi_1(t_{1k})$. Similarly, a particular trajectory in the subspace $\Phi_2(t_{2k})$ can be reconstructed from $y_2(k)$. This has been illustrated in Fig. 7., wherein $\Phi_1(t_{1k})$ and $\Phi_2(t_{2k})$ are capturing different observable sub spaces or projections of the process on to various manifolds observable to $y_1(k)$ and $y_2(k)$ respectively:

$$\Phi_1(t_{1k}) = g_1(y_1(k)) \tag{6}$$
$$\Phi_2(t_{2k}) = g_2(y_2(k))$$

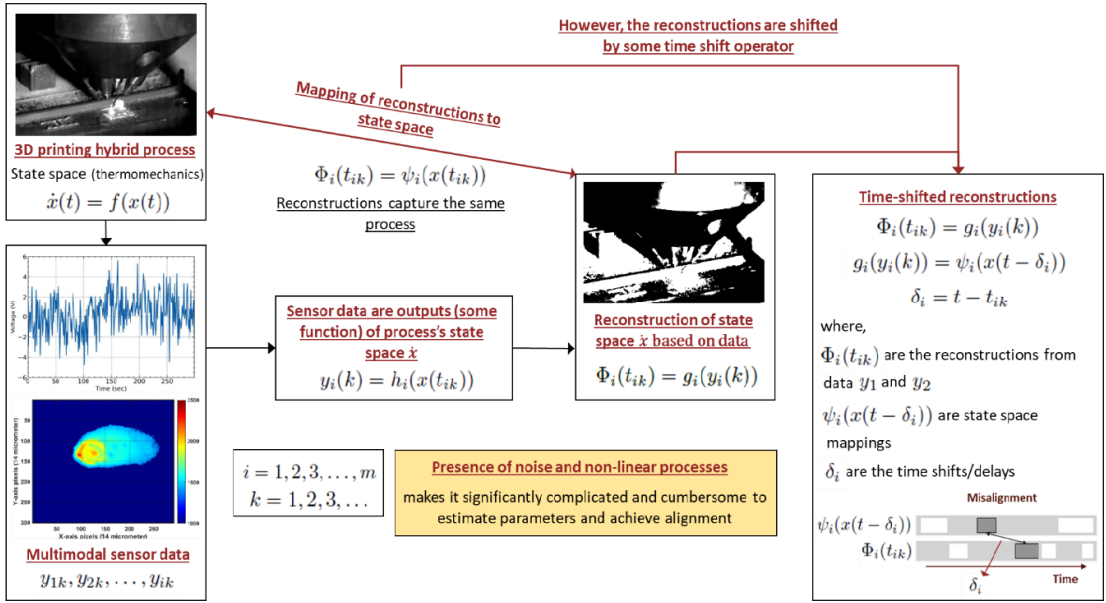

**Figure 6. Mathematical formulation and schema-based explanation of the alignment problem**

Here, $\Phi_1(t_{1k})$ is capturing a sub space of the state space only in the $x_1 x_3$ plane while $\Phi_2(t_{2k})$ has only been realized in the $x_1 x_2$ plane, both capturing different aspects of the process. However, since $\Phi_1(t_{1k}), \Phi_2(t_{2k}), \ldots, \Phi_i(t_{ik})$ are capturing the same process, they have a particular mapping relative to $\dot{x}$ such that:

$$\Phi_1(t_{1k}) = \psi_1(x(t_{1k})) \tag{7}$$
$$\Phi_2(t_{2k}) = \psi_2(x(t_{2k}))$$

and so on, reconstructed to actual state space mapping. These $\psi(.)$ functions are often time varying (non-autonomous). In the sense that, it usually has a time shift operator. In other words, the state space reconstructed from $y_1(k)$ i.e. $\Phi_1(t_{1k})$ might be time-shifted from the actual process by some unknown value $\delta_1$. Similarly, $\Phi_2(t_{2k})$ which comes from $y_2(k)$ could have shifted by some unknown value $\delta_2$ such that:

$$\Phi_i(t_{ik}) = g_i(y_i(k)) = \psi_i(x(t - \delta_i)) \tag{8}$$
$$\delta_i = t - t_{ik}$$

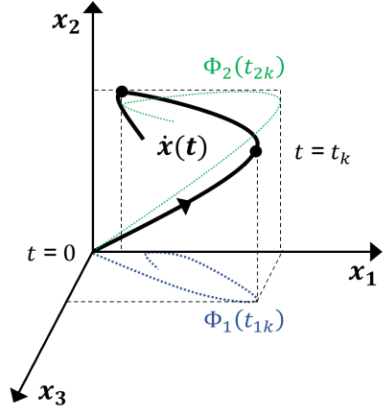

**Figure 7. Process trajectory and state subspace reconstructions**

The alignment problem essentially involves identifying those $\Phi_i$s and thereby those $\delta_i$s such that the outputs and the reconstructed state spaces align with the original state space. The multimodal data alignment problem is a challenging one and inherently shares similarities with the synchronization problem in nonlinear, dynamic, and chaotic systems [58]. Section 3 covered a formal treatment of this problem for a deterministic process. Even in cases where it is linear and autonomous, such alignment is mathematically difficult because it makes the determination of $\delta_i$s significantly complicated.

When the dynamics is nonlinear and the process has noise, the system operates in non-stationary regimes with different kinds of transience, such synchronization will not be possible due to the extremely cumbersome estimation of $\delta_i$s. A close form way of aligning various data outputs or developing a function that aligns them is almost impossible. Even in the absence of noise, if the process is non-linear, identification of these reconstructions as well as alignment becomes cumbersome. Existing methods have looked at addressing such alignment based on physical markers or statistical methods (entropy, correlation, etc.). However, they do not effectively work when a process involves a combination of nonlinear dynamics and presence of noise. Therefore, one must use domain knowledge, i.e., full understanding of the process to accomplish precise alignment. In the proposed work, we take advantage of the projections or process signature markers, and we go back to the process of deriving these in a more efficient way. The algorithms proposed are aimed at identifying those $\delta_i$s and the presented work is one of the first formal attempts at deriving a generalizable algorithmic solution for multimodal data alignment. The next section takes a deep dive into these algorithms corresponding to HiRA-Pro. Additional time-based multimodal data alignment algorithms are also proposed but are not the topic of priority and hence provided as a part of supplementary section.

## 4 Process signature-based multimodal data alignment (HiRA-Pro)

Process signature-based alignment algorithm involves two main steps - 1) identification of right process signatures, and 2) segmentizing or chunking various data streams based on identified process signatures to generate aligned multimodal data voxels. Process signatures are identifiable in various means depending on the type of data stream. They usually occur in the form of amplitude spikes, bursts, or are noticeable as changes in intensity values across the data. It is straightforward for machine log-based data wherein discrete event timestamps are usually captured for these identifiable signatures. These signatures must be identifiably present across all the data streams to align them using this methodology.

The proposed algorithm A0 is utilized for identifying and extracting a dictionary containing process physics and dynamics information in form of identifiable process signatures for each data stream $\varphi_i$. Depending on the data stream classification, extracting this information varies in the methodology. The proposed algorithm A0 takes in a set of $m$ data streams, $\{\varphi_1, \varphi_2, \varphi_3, \ldots, \varphi_m\}$ of dimensions $\{n_1, n_2, n_3, \ldots, n_m\}$ as input along with their classification to indicate whether they are time series-based $\{TS\}$, intensity-based $\{I\}$, or machine log-based $\{M\}$. It provides a dictionary $pd$ containing process signatures associated with each data stream $\varphi_i$ in the form $\{\{\pi_1^1, \pi_2^1, \pi_3^1, \ldots\}, \{\pi_1^2, \pi_2^2, \pi_3^2, \ldots\}, \ldots, \{\pi_1^m, \pi_2^m, \pi_3^m, \ldots\}\}$.

As the first step, it is important to know about the process which comes with experience in the field. Another alternative is to study the G-code associated with the process, but this is restricted to applications that involve machine movement based on some G-code. For time series-based data streams, the signals capture important events during the process in the form of spikes, bursts, or drops in the amplitude values (see Fig. 2.). By using these, it is possible to identify shift events such as the point at which the tool started and finished interacting with a workpiece or when the spindle started rotating. These depend a lot on the important concept of signal-to-noise ratio in order to capture such phenomena based on various settings of signal conditioning, gain, and amplification, among others.

---

**Algorithm A0 Process signature extraction**

**Inputs:**
- Set of data streams: $\{\varphi_1, \varphi_2, \varphi_3, \ldots, \varphi_m\}$, where $m$ is number of data streams with dimensions $\{n_1, n_2, n_3, \ldots, n_m\}$
- Data stream classes: time series-based $\{TS\}$, intensity-based $\{I\}$, and machine log-based $\{M\}$
- Baseline/Reference data stream index: $b$ s.t. $\varphi_b \leftarrow TS$

**Outputs:**
- $pd$: a dictionary containing prospective process physics and dynamics information for each data stream $\varphi_i$: $\{\{\pi_1^1, \pi_2^1, \pi_3^1, \ldots\}, \{\pi_1^2, \pi_2^2, \pi_3^2, \ldots\}, \ldots, \{\pi_1^m, \pi_2^m, \pi_3^m, \ldots\}\}$

1:     Extract and delineate process information based on domain knowledge (For example, G-code
2:     provides the machine tool path in machine language)
3:     **for** $i = 1$ to $m$ **do**
4:         **if** $\varphi_i \leftarrow TS$ **then**
5:             Preprocess the data (For example, demeaning signals)
6:             Extract actual sampling rate
7:             Splice or pad the data stream based on comparison with $\varphi_b$
8:             Find indices for amplitude spikes, bursts, or drops
9:         **else if** $\varphi_i \leftarrow I$ **then**
10:            Open $\varphi_i$ as a video file via *OpenCV* module:
11:            $vid = \text{cv2.VideoCapture}(\varphi_i)$
12:            Intensities $= [\ ]$
13:            **for** frame in $vid$ **do**
14:                 Find the intensity of the frame
15:                 Int $=$ frame.$mean(\ )$
16:                 Intensities.$append$(Int)
17:            Find indices for intensity amplitude spikes, bursts, or drops
18:         **else if** $\varphi_i \leftarrow M$ **then**
19:            Open $\varphi_i$ as dataframe via *pandas* module
20:            Find indices for discrete process events (For example, spindle turned on)
21:     **return** $\{\{\pi_1^1, \pi_2^1, \pi_3^1, \ldots\}, \{\pi_1^2, \pi_2^2, \pi_3^2, \ldots\}, \ldots, \{\pi_1^m, \pi_2^m, \pi_3^m, \ldots\}\}$

---

This is because not all phenomena result in visual changes in the signal of the same orders. For instance, turning the spindle on could result in a burst in amplitude values of a vibration sensor, but this burst will be significantly lower in amplitude as compared to periodic burst amplitudes when the tool is actually cutting the workpiece. Therefore, these settings need to be set appropriately. For example, it should be maintained at higher values in cases where not a lot of vibrations are generated. By doing so, we can amplify the signal-to-noise ratio and inherently also catch events such as printing tracks (interaction of 3D-printing powder with the baseplate) during a DED-based printing process. It is important to note that, even if the conditions are set to one setting, the sensors do capture a lot of process signatures associated with the process. But these captured events are just not prominently visible enough in the time domain, and further exploration of frequency domain and time-frequency domain are required. A lot of times, some sensor is lagging behind another in terms of acquisition, usually on a millisecond scale. Given that the timestamp itself does not matter in this alignment, one can move around the lagging signal such that it coincides with the leading one. The two main methodologies for doing so are slicing and padding, but depending on the stack used, the terminology or functions would vary. For example, Python provides this in the form of *insert()*, *append()*, and *extend()*, while MATLAB allows it in the form of simple addition of two arrays. In the case of intensity-based streams of high speed

---

**Algorithm A1 Process signature-based data alignment**

**Inputs:**
- Set of data streams: $\{\varphi_1, \varphi_2, \varphi_3, \ldots, \varphi_m\}$, where $m$ is number of data streams with dimensions $\{n_1, n_2, n_3, \ldots, n_m\}$
- Data stream classes: time series-based $\{TS\}$, intensity-based $\{I\}$, and machine log-based $\{M\}$
- $pd$: a dictionary containing prospective process physics and dynamics information for each data stream $\varphi_i$: $\{\{\pi_1^1, \pi_2^1, \pi_3^1, \ldots\}, \{\pi_1^2, \pi_2^2, \pi_3^2, \ldots\}, \ldots, \{\pi_1^m, \pi_2^m, \pi_3^m, \ldots\}\}$

**Outputs:**
- Set of aligned segmented datasets: $\{\{\xi_1^1, \xi_2^1, \xi_3^1, \ldots, \xi_p^1\}, \{\xi_1^2, \xi_2^2, \xi_3^2, \ldots, \xi_p^2\}, \ldots, \{\xi_1^m, \xi_2^m, \xi_3^m, \ldots, \xi_p^m\}\}$

1:     Find $p$ by examining $pd$ where $p$: prominent process signatures common to all data streams $\varphi_i$
2:     (For example, $TS$ amplitude spikes, bursts, or $I$ intensity changes across frames)
3:     **for** $i = 1$ to $m$ **do**
4:         **if** $\varphi_i \leftarrow TS$ **then**
5:             **for** $j$ in $p$ **do**
6:                 Find the indices for $j$ and $j+1$ in $\varphi_i$
7:                 $\xi_j^i = \varphi_i[idx_i(j): idx_i(j+1)]$
8:         **else if** $\varphi_i \leftarrow I$ **then**
9:             Open $\varphi_i$ as a video file via *OpenCV* module:
10:            $vid = $ cv2.VideoCapture($\varphi_i$)
11:            **for** $j$ in $p$ **do**
12:                Find the indices for $j$ and $j+1$ in $\varphi_i$
13:                $\xi_j^i = $ cv2.VideoWriter($\varphi_i$, video codec, desired FPS, $idx_i(j), idx_i(j+1)$)
14:         **else if** $\varphi_i \leftarrow M$ **then**
15:             Open $\varphi_i$ as dataframe via *pandas* module
16:             **for** $j$ in $p$ **do**
17:                Find the indices for $j$ and $j+1$ in $\varphi_i$
18:                $\xi_j^i = \varphi_i[idx_i(j): idx_i(j+1)]$
19:     **return** $\{\{\xi_1^1, \xi_2^1, \xi_3^1, \ldots, \xi_p^1\}, \{\xi_1^2, \xi_2^2, \xi_3^2, \ldots, \xi_p^2\}, \ldots, \{\xi_1^m, \xi_2^m, \xi_3^m, \ldots, \xi_p^m\}\}$

---

camera and optical camera captures, the imaging sensors capture information about the process that is hidden in the frames (see Fig. 4.). With modern Python packages such as *OpenCV*, it is possible to extract derivatives of these frames in the form of changes in the intensity magnitudes across the frames. For generating derivatives, either summation of all the pixel intensities is carried out or simply averaged over to generate a set of mean intensity values against frame count. It is possible to capture easy-to-identify events like a laser turning on or off, as well as other complex events such as a spindle going up and down or a tool interacting with the workpiece as well. This is done by mimicking the process followed in case of time series-based data streams by monitoring intensity amplitudes instead of forces or vibrations. Most machine logs capture discrete machine events, and it is easy to extract process signatures directly from these logs (see Fig. 5.) for the case of machine log-based data streams. The process signatures dictionary is generated based on these captured process signatures and their corresponding start and end indices for each of data streams in the multimodal data.

Once the process signatures dictionary, $\{\{\pi_1^1, \pi_2^1, \pi_3^1, ...\}, \{\pi_1^2, \pi_2^2, \pi_3^2, ...\}, ..., \{\pi_1^m, \pi_2^m, \pi_3^m, ...\}\}$, is available, process signature-based multimodal data alignment is achieved by first finding prominent process signatures $p$ that are common to all the data streams $\varphi_i$. These signatures capture the part of the process that are prominently differentiable from the rest of the data for each of the data streams. By following algorithm A1 and based on dictionary output from algorithm A0, segmentized sections of the process multimodal data are extracted based on indices of prominent process signatures that are common to all data streams $\varphi_i$. It is straightforward for time series-based and machine log-based data streams but require reading frames between start and end indices and rewriting smaller segments of the video captures for intensity-based ones. This also provides for choosing any alternative FPS and other relevant settings as desired. Algorithm A1 results in a set of aligned segmented datasets: $\{\{\xi_1^1, \xi_2^1, \xi_3^1, ..., \xi_p^1\}, \{\xi_1^2, \xi_2^2, \xi_3^2, ..., \xi_p^2\}, ..., \{\xi_1^m, \xi_2^m, \xi_3^m, ..., \xi_p^m\}\}$ where $p$ represents the prominent process signatures common to all data streams $\varphi_i$, $m$ is the number of data streams in total, and each $\{\xi_1^i, \xi_2^i, \xi_3^i, ..., \xi_p^i\}$ represents aligned and segmented voxels of a data stream $\varphi_i$.

Benchmarking and testing the process signature-based multimodal data alignment on various data streams indicate results with sub-millisecond and even microsecond resolutions. However, these are difficult to show quantitative as compared to the case of time-based multimodal data alignment since no timestamps are involved as such. Like earlier, the results may vary depending on factors such as acquisition rates of various data streams. As a significant benefit, such process physics-driven alignment does not rely on the time information available with various streams, effectively getting rid of any challenges that come from unsynchronized clocks, their maintenance, and variability in their associated formats. These are safer in terms of any cyber-attacks and more applicable to a wide range of practical settings, even when some data stream acquisitions are proprietary. Against timestamps, the prominent process signatures always represent a particular event that took place accurately and there is no digitization and other inconsistencies as such. Such process signature-based synchronization, alignment, and voxelization of multimodal sensors is essential for any smart manufacturing implementation to be at forefronts of industry 4.0 due to its ability to provide a more accurate representation of a manufacturing process.

## 5 Experimentation and case study demonstrations

### 5.1 Experimental setup

The Texas A&M smart manufacturing implementation consists of an Optomec-LENS® MTS 500 hybrid machine which is capable of both additive and in-situ finishing operations. The machine tool is essentially a 4-axis CNC machine that automatically controls the worktable motion in $X$ and $Y$ directions, motion of the laser and a milling head along a $Z$ axis, and the rotation of a horizontal spindle, which can be used to clad and repair a variety of freeform parts. The machine, through a Siemens 828D controller, provides control over multiple process parameters such as powder composition, laser power, and dwell time. The associated controller variables in a process are always tracked via an OPC UA data logger wherein several variables such as the table motion along $X$, $Y$, and $Z$ directions, RPM of the spindle, etc., can be continuously monitored and saved as a log. The experiments consisted of printing stainless steel 316L samples of dimensions $10 \times 10 \times 10$ mm$^3$, followed by milling them with sub micrometer precisions via an end mill tool of $0.375$″ diameter.

The machine has been sensorized by instrumenting various COTS and image-based sensors. In total, the implementation is capable of simultaneously collecting 13+ different data streams throughout a process. The implementation also hosts videographic capabilities and the entire process is captured via an optical camera, along with a Stratonics ThermaViz melt pool sensor capturing the thermal variations during printing. It also consists of a Photron Mini AX200 high speed camera that allows observation of milli- or micro-second phenomena at up to 960000 frames per second. The schematic of the smart manufacturing implementation is illustrated in Fig. 8. The next two sub-sections within this section cover a specific and a broader case study respectively to demonstrate the applicative power of HiRA-Pro multimodal data alignment methodology.

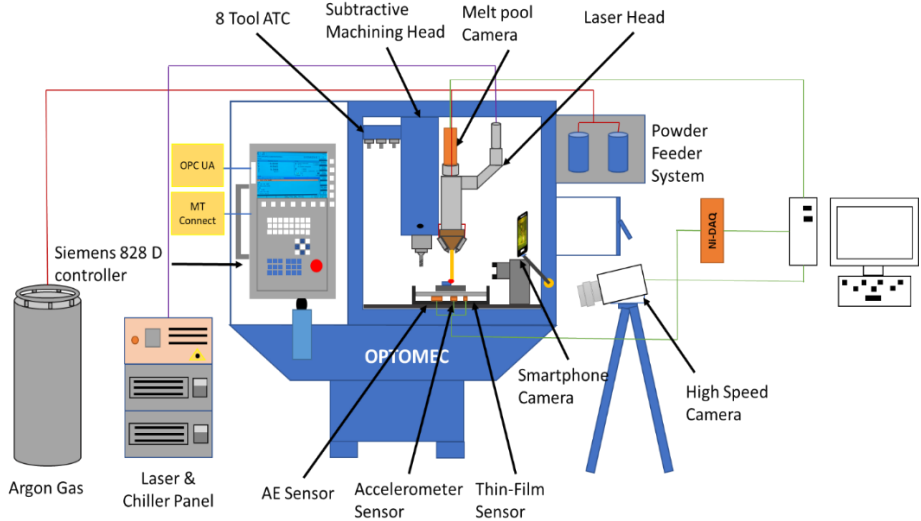

**Figure 8. Optomec-LENS® MTS 500 hybrid machine schematic**

## 5.2 Specific case study: HiRA-Pro enhanced porosity prediction

In this specific case study, the power of HiRA-Pro has been demonstrated in achieving improved ML-based predictive performance for detecting porosity on a voxel-level on a manufacturer part using hybrid DED process [59]. HiRA-Pro is employed for modalities based on an accelerometer, an AE sensor, and a thermal meltpool sensor. High spatio-temporal resolutions of 0.5 mm and < 1 ms were achieved, allowing for accurate porosity prediction in surface elements measuring $1 \times 1$ mm$^2$, known as surfels (2D voxels on the top surface). Due to multimodal data alignment and synchronization, the integrity of information flow was maintained, correlating it precisely with the manufacturing process of each surfel.

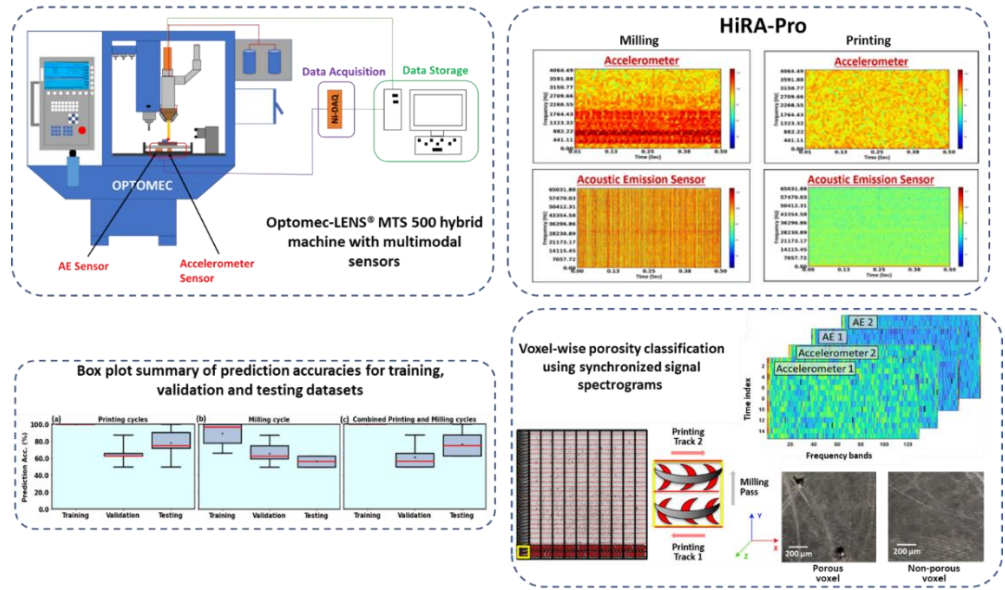

**Figure 9. Porosity prediction case study graphical illustration**

HiRA-Pro addressed the challenge posed by varying sampling rates and modalities of the thermal, accelerometer, and AE data, that were acquired at 30, 10000, and 100000 Hz, respectively. Thereby, this ensured that time-frequency patterns across different modalities were accurately aligned, allowing for properly tensorizable training and testing data for further ML-based modeling efforts. Due to the alignment, the classification test accuracies improved from around 65% to an impressive 87.5% even with limited (∼80) voxels. Figure 9 illustrates this porosity prediction case study in a graphical abstract manner, only providing necessary amounts of details. While HiRA-Pro was employed on three modalities, only two, namely - accelerometer and AE sensor, are driving the porosity prediction. The thermal meltpool sensor is later used to quantify existence of spatter events and successfully correlated to the rest of the findings. The findings imply that there is a clear distinction of energy levels between spatter events in porous surfels vs spatter events in non-porous surfels.

The findings underscore the effectiveness of HiRA-Pro in enhancing the quality of porosity prediction in additive manufacturing processes. With a very high testing classification accuracy of detecting porosity within surfels, the aligned data from the printing cycles proved to be particularly sensitive to such occurrences. This highlights the critical role of accurate multimodal data alignment in enabling precise predictive analytics, offering valuable insights for future research and applications in improving manufacturing quality control.

### 5.3 Broader case study: HiRA-Pro for hybrid 3D-printing operation

The process signature-based as well as time-based (see supplemental material section S.1) multimodal data alignment methodology demonstrated excellent capabilities for achieving high resolution time-space alignment for experiments (hybrid 3D-printing operations) carried out on an Optomec-LENS® MTS 500 hybrid machine. The experiments involved 3 time series-based data streams, namely those collected via a thin-film sensor (TF), an accelerometer (Acc), and an acoustic emission (AE) sensor that were collected at sampling rates of 1024, 8192, and 131062 Hz respectively in NI TDMS (technical data management system) file format via LabView throughout the duration of experiment that lasted 25 minutes. There was a total of 3 intensity-based data streams in form of high-speed camera (HSC) video capture captured at 6400 FPS via Photron Mini ax200, optical camera (smartphone) (OC) recording at 60 FPS, and a Stratonics ThermaViz melt pool capture (TVM) at 60 Hz. For the machine log-based data stream, OPC data logger (OPC) connected with a Siemens 828D controller was responsible for controlling and capturing variables pertaining to table motions along X, Y, Z directions and RPM of spindle, among others at around 20-30 Hz.

Time-based multimodal data alignment (see supplemental material section S.1) alone resulted in a best resolution of 0.001 seconds with an average resolution of 0.004 seconds among various data streams and the least resolution of 0.290 seconds, outclassing existing alignment methodologies by at least an order of magnitude. Table 1 captures this in the form of average timestamp difference between start and end times of aligned data streams. The least resolution is between TVM and OPC data streams while the best resolution is obtained for the case of Acc and TF data streams. Upon further analyses, it was found that the least resolution resulted in a larger number due to varying ranges of sampling rates. The AE data stream was acquired at a very high sampling rate while the slowest was only around 20 Hz. This can be easily solved and verified by redoing the time-based alignment with up-sampled slower data streams and reinforcing the slower signals. Realignment indicates similar resolutions in the range of 1 millisecond or so.

**Table 1. Time-based multimodal data alignment results (in seconds) on Optomec-LENS® MTS 500 hybrid machine by considering Thin-film sensor (TF) as baseline data stream**

| Data Stream | TF | Acc | AE | OPC | TVM | HSC | OC |
|---|---|---|---|---|---|---|---|
| TF | 0 | 0.001 | 0.005 | -0.209 | 0.081 | 0.051 | -0.044 |
| Acc | | 0 | 0.004 | -0.21 | 0.08 | 0.05 | -0.045 |
| AE | | | 0 | -0.214 | 0.076 | 0.046 | -0.049 |
| OPC | | | | 0 | 0.29 | 0.26 | 0.164 |
| TVM | | | | | 0 | -0.03 | -0.125 |
| HSC | | | | | | 0 | -0.095 |
| OC | | | | | | | 0 |

Process signature-based multimodal data alignment results in sub-millisecond resolutions of alignment. For the case study considered, the list of process signatures that were considered for achieving multimodal data alignment is indicated in Table 2. These include process signatures for both printing and milling operations involved in the experiments which were carried out on the hybrid machine. Table 2 also illustrates the type of activity that is picked up for various process signatures by various data streams. For example, turning the laser on results in an intensity spike in intensity-based data streams OC, HSC, and TVM while turning the laser off is captured in form of an intensity drop. It is important to note that TVM does not play a role in the case of milling operation since there is no laser head involved. Therefore, TVM data stream will not capture any temperature variations for milling. Apart from the process signatures listed in Table 2, it is possible to use several others depending on the operation being captured. For instance, with proper conditioning and amplification settings, milling tool interactions with a part can be captured via time series-based data streams such as Acc (see Fig. 2). Time-based alignment depends primarily on the sampling rates of various data streams. If all the streams operate at very high sampling rates such as the AE data stream and synchronized time information is available throughout, high resolution alignment among all the data streams is possible.

Table 2. Process signatures dictionary for process signature-based alignment

| Process Signature | TF, Acc, AE | OPC | OC, HSC | TVM |
|---|---|---|---|---|
| **Printing** | | | | |
| Laser on and off | | | Intensity spikes and drops | |
| Layer starts | | Z position | Intensity spikes | |
| Layer perimeter rapid traverses | Amplitude spikes | XY positions | Intensity drops | |
| Layer track rapid traverses | | | | |
| Layer ends | | Z position | | |
| Layer end rapid traverses | Amplitude spikes | XYZ positions | | |
| Upward and downward motions of printing head | | Z position | Intensity variations | |
| **Milling** | | | | |
| Spindle on and off | Amplitude spikes | Spindle RPM | Intensity variations | Not Applicable |
| Pass starts | | XYZ positions | | |
| Pass ends | | | | |
| Pass end rapid traverses | | | | |
| Upward and download motions of milling tool | | Z position | | |

In the case of process signatures, a majority of process signatures such as the impact events of start and stop of a component in a process do not get captured in the form of a single peak or impulse, but instead are captured as waveforms. Such waveforms will vary in width depending on the duration of the impulse or signature being captured. For instance, the "laser turn on" event is a very short duration impulse in matter of fractions of a second, resulting in around a few thousand points wide waveform on an AE signal. On the other hand, a rapid traverse is a medium duration impulse that takes places in several hundreds of milliseconds, and it results in a much wider waveform. The alignment precision and resolution would be proportional to the width of the waveform. The alignment resolution therefore varies based on the chosen process signature.

If identified signature is an impulse such as table retraction, it will result in millisecond resolution, while if it is something like a tool or tooth interaction during a milling pass, it can result in sub-millisecond or even microsecond resolutions. Figure 9 depicts such a signature in the form of a waveform corresponding to a milling tool's tooth interaction with a 3D-printed part. Considering tooth interaction 4, the waveform starts at around index of 3542 and ends at approximately 3544.5 milliseconds. This is equivalent to approximately 20-25 samples wide waveform, representing the process signature of a tooth interacting with the part captured via an accelerometer. The worst-case alignment resolution in such a case would be 20/8192 ≈ 0.002 to 25/8192 ≈ 0.003 seconds.

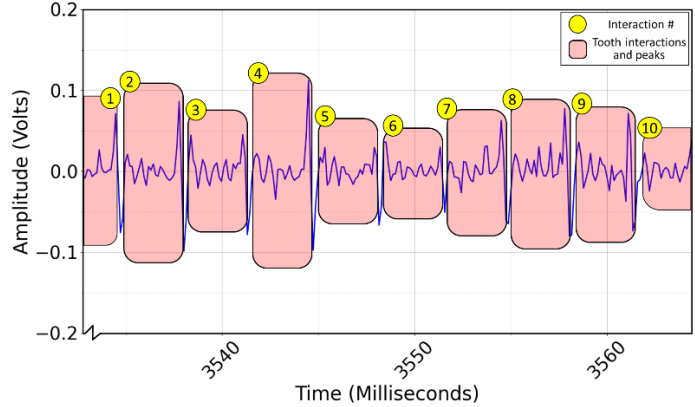

**Figure 9. Example process signature as a waveform**

The hardware-based or physical trigger-based alignment considers the entire waveform (internal triggers) while triggering, or even worse the time stamp (external triggers) of a particular event. So, such methods result in time resolutions that correspond to at least the entire width of the waveform. However, in the proposed process signature-based multimodal alignment, we only use specific parts of the wave to achieve desired resolution of alignment, namely, central most peaks in waveforms are used across a data stream to capture various process signature stamps. This is repeated for all the data streams and the common signatures are then used for aligning the data streams together. Additionally, short time impulses such as laser turning on can result in resolutions in the range of ≤ 0.001 seconds.

**Table 3. Alignment methodology resolution comparison**

| Alignment methodology | Width of the Waveform | Resolutions (in seconds) |
|---|---|---|
| External triggering | ≥ Highest sampling rate | ≥ 1 |
| Internal triggering | ~$\frac{1}{1000}$ to $\frac{1}{10}$ of Highest sampling rate | ~0.001 to 0.1 |
| Time-based | | |
| Process signature-based | ~$\frac{1}{10000}$ to $\frac{1}{1000}$ of Highest sampling rate | ≤ 0.001 |

Table 3 provides a comparative study of possible resolutions based on various multimodal alignment methodologies, demonstrating the power of the proposed method. While trigger-based methods can provide resolutions of several hundreds of milliseconds, the time-based alignment (see supplementary material section S.1) provides resolutions in the range of couple of milliseconds and the process signature-based alignment can result in sub-millisecond resolutions. Figures 10(a) and (b) provide illustration of aligned multimodal data streams that involve spectrogram videos for time series-based data streams, a visualization of 3 channels of machine log-based data streams, and three standalone intensity-based data streams for milling and printing operations respectively. In Fig. 10(a), the thermogram captured in the thermal meltpool data stream remains at lower temperatures (dark blue on the color bar) consistency since no laser is involved during milling operation.

Process signature-based multimodal data alignment allows for involvement of the physical process or part itself, as well as metrological aspects of the part, resulting in high resolution time-space alignment and registration. This is done so by identifying the right set of voxels from the aligned multimodal data. Figure 11 illustrates this in the form of a stitched optical microscopy capture of the topmost layer of the printed part. Both printing and milling operation directions are upwards as indicated in Fig. 11. By additional post-processing of the associated meta-data, it can be decomposed back into multiple smaller captures that register back to voxels of aligned multimodal data. This aspect of stitching and then decomposing is important in such metrological data because the resolution of the equipment capturing these, and the desired voxel dimensions might differ.

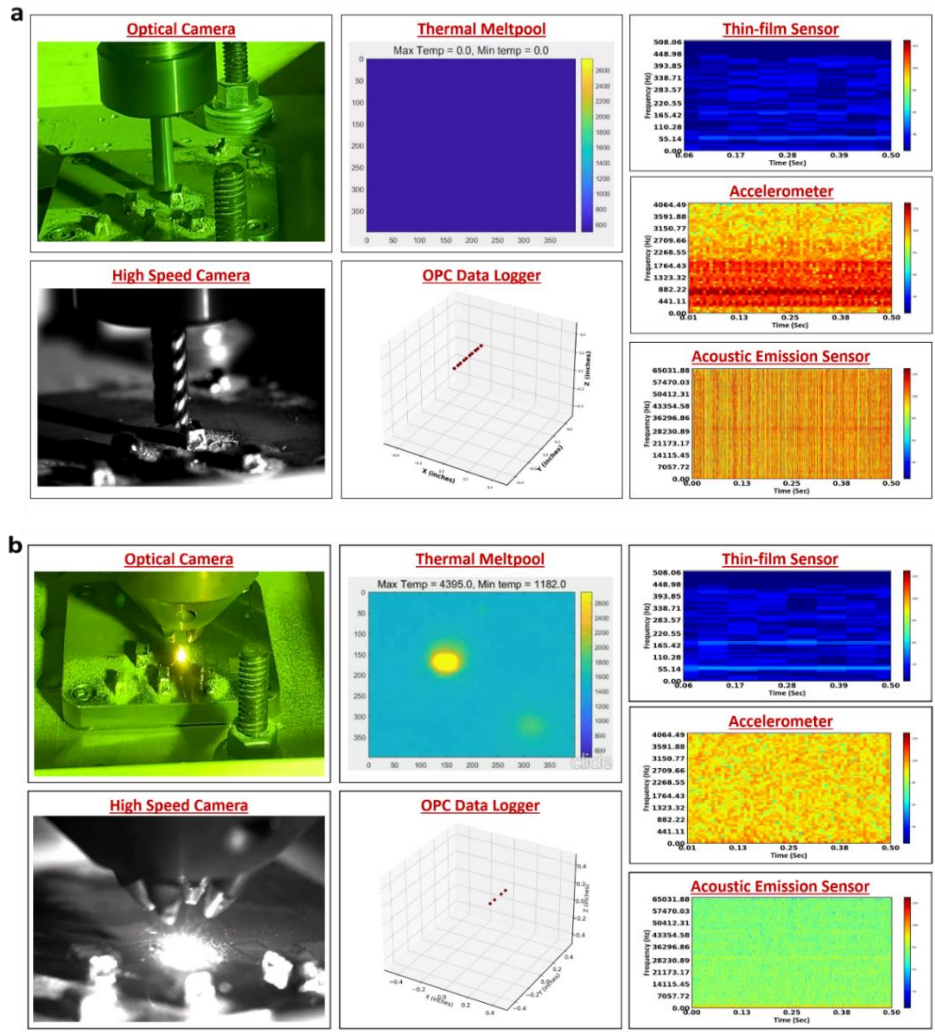

**Figure 10. Process signature-based multimodal data alignment results for a) Milling cycle, b) Printing operation**

Table 4 provides a comparison among the proposed process signature-based alignment, the supplementary time-based methodology, and hardware trigger-based alignment methods by covering benefits and disadvantages of each of them. While hardware trigger-based methods have specific hardware requirements such as SLS-based acquisition systems, the two proposed methodologies do not require any additional hardware. Another advantage of the proposed methods lies in the fact that they both operate as post-processing operations and are viable for previously collected experimental datasets as well, making them much more robust and generalizable in nature.

Process signature-based and time-based alignment results in sub-millisecond and millisecond resolutions, outperforming the hardware trigger-based methods that usually employ the entire width or timestamp of a particular signature for achieving roughly aligned data acquisition. Voxelization or data segmentation corresponding to various event groups in a process is an upcoming research aspect that is much more flexible via process signature-based multimodal alignment that captures the process signature information and provides for a voxel-wise alignment.

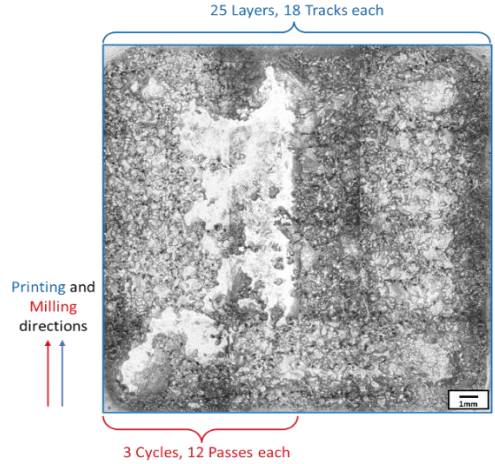

**Figure 11. Metrological aspect of the printed part in the form of a stitched optical microscope capture of the top surface**

**Table 4. Comparison of hardware-, time-, and process signature-based multimodal data alignment**

| Hardware trigger-based | Time-based | Process signature-based |
|---|---|---|
| Specific hardware required (for example: SLS) | No hardware requirements | |
| Alignment resolution in several hundreds of $ms$ | Alignment resolution of up to $1 ms$ | Estimated alignment resolutions in sub-milliseconds and up to $1 \mu s$ |
| Requires additional steps for data voxelization | | Voxelization (creating voxels) is more accurate and easier to generate |
| Difficult to isolate defective events | | Easier to identify and localize events such as spatter during printing, chip breakage during milling, etc. |

The high-resolution spatio-temporal alignment provided by HiRA-Pro also allows for isolation of very short duration events such as spatter during printing and chip breakage during milling, making ML and AI model predictions powerful by using such aligned and voxelized data. Such alignment also allows for variations in desired resolutions depending on the choice of events. For instance, it is very straightforward to obtain events such as engagement of the tool with the work piece via machine log-based event logs, but alignment based on this might suffer due to the low sampling rate (~20 Hz) and result in low resolution alignment. The same event when identified via time domain signals of accelerometer or acoustic emission sensor can provide very high sampling rate availability, resulting in a highly precise stamping of the event. However, this can suffer from high computational requirements due to large amounts of data and its post processing.

In certain channels, clock delays are common in the form of DC drifts (fixed delay) and variable delays during acquisition. Time-based multimodal data alignment is needed to deal with these delays in some ways. Algorithm S1 (see supplementary material section S.1) provides a potential solution for such cases, but handling the variable delay might be challenging in certain cases. In such scenarios, it is advisable to go for process signature-based alignment which allows for highly precise process mechanics-based alignment across multiple channels of data without relying on any absolute time information.

# 6 Conclusions and future work

Data alignment is a central problem when dealing with multimodal data. The alignment precisions determine the resolution with which one can monitor and control a process in time as well as the quality (e.g., geometry, microstructure, porosity, etc.) of a product in space. However, not much attention has been given to how precisely one can align data streams in real-time. Conventional approaches depend purely on clock synchronization, statistical correlations, and common external triggering. They often lead to erroneous and imprecise alignment, especially for identifying anomalies at high temporal and spatial resolutions. The proposed work is perhaps the first attempt to demonstrate the use of process signatures to enhance the precisions of alignment of multimodal data. Process signature-based multimodal data alignment can result in highly precise alignment with millisecond and sub-millisecond resolutions. The proposed method results in aligned data in a voxelized manner, further paving the way for accurately registering the data with the actual process, part, or any associated metrological scans.

We envision the following as future work in this untapped area of research based around multimodal data alignment. In the proposed work, the alignment is process physics-driven, requiring some understanding of the physical process and linking it with the data streams. To address this, the Texas A&M team is currently involved in pitching the proposed alignment methodology into the market as a generalizable and scalable end-to-end solution, in turn minimizing the domain expertise to some extent. Another interesting area to explore would involve adding in a layer of uncertainty quantification, especially when dealing with complex, stochastic, and dynamic processes. A rigorous quantification and communication of uncertainty in such alignment further enhance the reliability of alignment results, providing a better understanding of confidence levels associated with the aligned data and facilitating informed decision-making in uncertain environments. Given the algorithmic nature, reducing the associated time and space complexity would be an interesting follow-up work. Additionally, algorithmizing the data registration process in terms of substantiating metrological micrographs such as profilometer and microscopic images with the aligned and voxelized data, would be a viable future work. Furthermore, with the onset of 5G and increasing cyberthreats, security, and privacy concerns, potential future work also involves benchmarking the proposed multimodal data alignment methodology with existing works and unaligned datasets in terms of the security. It would be interesting to quantify the security implications as a comparative study between unaligned raw data against aligned data derived from such process signature-based alignment. The issues related to data privacy, bias, and fairness can also be evaluated. Nonetheless, this work resides in a largely unexplored, yet critical intersection of data science and process physics. Its potential to improve the quality of data and decision-making, coupled with inherent challenges, should serve as a strong motivation for future innovation in this exciting domain of research.

## Acknowledgments

This work was supported by the Rockwell International Professorship and United States National Science Foundation (NSF) under ECCS-1953694 and IIS-1849085.

# Supplemental Material

The supplemental material for this paper covers description and algorithm details about a time-based multimodal data alignment methodology, which provides for a highly viable and scalable algorithmic route of achieving millisecond and sub-millimeter scale temporal-spatial resolutions.

## S.1 Time-based multimodal data alignment

The approach consists of two main steps derived in the form of two algorithms wherein one is used for extracting timestamps associated with data streams and other for synchronizing the extracted timestamps or the clocks of various data streams. While most modern data acquisition suites provide timestamp information, aligning data streams based on their timestamps can result in bizarre synchronization because there is no guarantee that all data source clocks are in-sync in the first place. One protocol allows for pre-synchronization of these individual clocks before any data acquisition and then extract time information associated with various channels by following algorithm S0 proposed in this work. However, such a protocol can quickly become tedious as the number of separate data channels being collected increases. Moreover, pre-synchronization of clocks do not necessarily last forever. Due to the inherent nature of the systems handling data acquisition, clocks quickly go out of sync, worsening it every time due to randomness. So, in this paper, we also propose a supplemental algorithm that allows for a post-synchronization of extracted timestamps of data streams. By doing so, the randomness of unsynchronous data acquisition for every iteration of data collection can be minimized significantly. The protocol for the post-synchronization does so by deriving formulae that remain consistent across various data acquisition iterations. The proposed work provides this in the form of supplemental algorithm S1.

---

**Algorithm S0 Data stream timestamps extraction**

**Inputs:**
- Set of data streams: $\{\varphi_1, \varphi_2, \varphi_3, \ldots, \varphi_m\}$, where $m$ is number of data streams with dimensions $\{n_1, n_2, n_3, \ldots, n_m\}$
- Data stream classes: time series-based $\{TS\}$, intensity-based $\{I\}$, and machine log-based $\{M\}$

**Outputs:**
- Set of corresponding clock timestamps: $\{t_1, t_2, t_3, \ldots, t_m\}$ in local time format with dimensions $\{n_1, n_2, n_3, \ldots, n_m\}$

1:    **for** $i = 1$ to $m$ **do**
2:        **if** $\varphi_i \leftarrow TS$ **then**
3:            Extract the clock timestamps $\{t_i\}$ in local time format via *nptdms* module:
4:            $t_i = \varphi_i.\text{time\_track}(absolution\_time = True)$
5:        **else if** $\varphi_i \leftarrow I$ **then**
6:            Open $\varphi_i$ as a video file via *OpenCV* module:
7:            $vid = \text{cv2.VideoCapture}(\varphi_i)$
8:            **if** $vid.\text{metadata} = True$ **then**
9:                Extract the clock timestamps $\{t_i\}$ in local time format via *OpenCV* module:
10:                $t_i = vid.\text{get}(\text{cv2.CAP\_PROP\_POS\_MSEC})$
11:            **else then**
12:                Apply *OCR* to extract clock timestamps $\{t_i\}$ in local time format via *pytesseract*
13:                module:
14:                $t_i = \text{pytesseract.image}_t\text{o}_s\text{tring}(vid)$
15:        **else if** $\varphi_i \leftarrow M$ **then**
16:            Extract the clock timestamps $\{t_i\}$ column in the local time format via *pandas* module
17:            $t_i = \text{pandas.read\_csv}(\varphi_i, \text{usecols} = [\text{'timestamps'}])$
18:    **return** $\{t_1, t_2, t_3, \ldots, t_m\}$

Based on the data stream classification, the steps involved vary as demonstrated in algorithm S0. Benchmarking and testing the algorithm on various data streams indicate results that correspond to time-based alignment with millisecond resolutions. The results can vary depending on factors such as acquisition rates of various data streams and selection of baseline for clock synchronization.

---

**Algorithm S1: Clock synchronization for data streams**

**Inputs:**
- Set of data streams: $\{\varphi_1, \varphi_2, \varphi_3, \ldots, \varphi_m\}$, where $m$ is number of data streams
- Set of corresponding clock timestamps: $\{t_1, t_2, t_3, \ldots, t_m\}$ in local time format with dimensions $\{n_1, n_2, n_3, \ldots, n_m\}$
- Baseline/Reference data stream index: $b$ s.t. $1 \leq b \leq m$

**Outputs:**
- Set of synchronized clock timestamps: $\{\tau_1, \tau_2, \tau_3, \ldots, \tau_m\}$ in UTC time format (For example, $yyyy\text{-}MM\text{-}dd\ HH\text{:}mm\text{:}ss.fff$)
- Set of synchronized clock timestamps: $\{t'_1, t'_2, t'_3, \ldots, t'_m\}$ in local time format
- Set of clock start times: $\{s_1, s_2, s_3, \ldots, s_m\}$
- Set of clock delays: $\{d_1, d_2, d_3, \ldots, d_m\}$

1:    Specify desired UTC time format for clock timestamps as a string $f$
2:    Extract clock start time for baseline data stream:
3:    $s_b = t_b[0]$
4:    **for** $i = 1$ to $m$ **do**
5:        Extract clock start time $\{s_i\}$ for data stream clock timestamps $\{t_i\}$:
6:        $s_i = t_i[0]$
7:        Calculate clock start time delay $\{d_i\}$ using clock start time $\{s_i\}$ and baseline data stream
8:        clock start time $\{s_b\}$:
9:        $d_i = (s_i - s_b).\text{total\_seconds}()$
10:       Check if data stream clock is leading or lagging compared to baseline data stream clock:
11:       **if** $d_i \geq 0$ **then**
12:           Synchronized clock timestamps $\{t'_i\}$ in local time format are calculated by subtracting
13:           clock start time delay $\{d_i\}$ in seconds from clock timestamps $\{t_i\}$:
14:           $t'_i = t_i - \text{timedelta}(\text{seconds} = d_i)$
15:       **else if** $d_i < 0$ **then**
16:           Synchronized clock timestamps $\{t'_i\}$ in local time format are calculated by adding clock
17:           start time delay $\{d_i\}$ in seconds to clock timestamps $\{t_i\}$:
18:           $t'_i = t_i + \text{timedelta}(\text{seconds} = d_i)$
19:       Convert synchronized clock timestamps $\{t'_i\}$ to specified UTC time format
20:       **for** $j = 1$ to $n_i$ **do**
21:           $\tau_i[j] = t'_i[j].\text{replace}(\text{tzinfo}=\text{timestamp.utc}).\text{strftime}(f)$
22:    **return** $\{\tau_1, \tau_2, \tau_3, \ldots, \tau_m\}, \{t'_1, t'_2, t'_3, \ldots, t'_m\}, \{s_1, s_2, s_3, \ldots, s_m\}, \{d_1, d_2, d_3, \ldots, d_m\}$

---

The proposed supplemental algorithm S0 for data stream timestamps extraction takes in a set of $m$ data streams, $\{\varphi_1, \varphi_2, \varphi_3, \ldots, \varphi_m\}$ of dimensions $\{n_1, n_2, n_3, \ldots, n_m\}$ as input along with their classification to indicate whether they are time series-based $\{TS\}$, intensity-based $\{I\}$, or machine log-based $\{M\}$. For each data stream $\varphi_i$, the algorithm first checks its classification. If $\varphi_i$ is a time series-based data stream ($\varphi_i \leftarrow TS$), the timestamps are extracted in local time format by using Python's *nptdms* module. To enable availability of this timestamp information, it is important to enable it while setting up the LabView-based data acquisition by adding a virtual clock in the architecture. If the acquisition of such signals is not via LabView, it is still possible to extract the time information by using suitable wrappers such as the *NI-DAQmx* for NI data acquisition and conditioning devices. If $\varphi_i$ is an intensity-based data stream ($\varphi_i \leftarrow I$), the extraction of timestamps require use of accessing associated meta-data or directly extracting timestamp information that is usually available on the frame via optical character recognition (OCR) algorithms. For both, Python's *OpenCV* module was used to read the data stream frame-by-frame. For extracting timestamps via metadata, video capture properties were used to identify every frame's position in millisecond. These frame timestamps can then be converted into proper format through some postprocessing. In the case of OCR, Python's *pytesseract* module was utilized to directly read the information available on the screen. To extract only the timestamp, the frames can be cropped to the region containing the time information and then OCR can be applied on a frame-by-frame basis. If $\varphi_i$ is a machine log-based data stream ($\varphi_i \leftarrow M$), the Python's *pandas* module can be used to directly extract the timestamps that are usually available in such machine logs in form of time information for various events. The algorithm returns the set of $m$ timestamps $\{t_1, t_2, t_3, …, t_m\}$ in local time format with dimensions $\{n_1, n_2, n_3, …, n_m\}$ corresponding to the $m$ data streams, $\{\varphi_1, \varphi_2, \varphi_3, …, \varphi_m\}$.

Once the timestamps $\{t_1, t_2, t_3, …, t_m\}$ in local time format corresponding to the set of $m$ data streams $\{\varphi_1, \varphi_2, \varphi_3, …, \varphi_m\}$ are available, synchronization is required. The proposed supplemental algorithm S1 takes in these timestamps along with a specified baseline or reference data stream among them as input. The desired time format, usually UTC format, for the timestamps is also taken as input to achieve formatting consistency. Given the myriad of sensing technology options that can operate at various contrasting sampling rates, some acquisition can be as slow as a few hundred Hz (imaging systems) while others can be several hundred kHz (time series-based sensors). In such cases, the selection of the baseline data stream is important and usually preferred to be the slowest acquisition sensor stream. As the first step of the algorithm, the clock start time for the baseline data stream $\varphi_b$ is extracted by taking the first timestamp $s_b$ from $t_b$. Followed by this, for every data stream $\varphi_i$, the algorithm compares the first timestamp of $t_i$, $s_i$ with $s_b$ to calculate the start time delay $d_i$ between various data streams and baseline. If $d_i \geq 0$, the data stream $\varphi_i$ is leading ahead of $\varphi_b$ in terms of timestamps. On the other hand, if $d_i < 0$, the data stream $\varphi_i$ is lagging behind $\varphi_b$ in terms of timestamps. Based on these delays, the algorithm either adds or subtracts the delay (in seconds) from $t_i$ to come up with synchronized timestamps $\{t'_1, t'_2, t'_3, …, t'_m\}$ for the $m$ data streams $\{\varphi_1, \varphi_2, \varphi_3, …, \varphi_m\}$. As the final step, using the specified time format, all the synchronized timestamps are converted from local time format to a unified and consistent format, resulting in a set of synchronized clock timestamps $\{\tau_1, \tau_2, \tau_3, …, \tau_m\}$ in UTC time format (For example, $yyyy\text{-}MM\text{-}dd\ HH:mm:ss.fff$).

Both time-based and process signature-based multimodal data alignment can provide excellent alignment with millisecond and microsecond resolutions. As discussed, time-based alignment is not always possible and good enough, and process signature-based alignment is difficult to quantify in terms of resolution numbers. Thereby, the proposed work also allows for a combination of algorithms S0 and S1 with A0 and A1. In such a combined multimodal data alignment process, while it is not required, it is advisable for clocks corresponding to all time series-based data streams to be in-sync and format. Other formatting consistencies such as timestamp formats and postprocessing are also required. For example, one sensor might be outputting UTC timestamps in the format, 2023-03-27T13:54:21Z, while other might be relying on ISO format, 2023-03-27T13:54:21+00:00. Both of these correspond to the same time and date but differ in their format. It is therefore desirable to bring all timestamps to a common format, ideally some UTC format.